\documentclass[aps,apl,reprint,amsmath,amssymb,superscriptaddress]{revtex4-1}
\usepackage[dvips]{graphicx}
\usepackage{graphicx}
\usepackage{dcolumn} 
\usepackage{graphics}
\usepackage{bbm}
\usepackage{amsmath,amsfonts,amssymb}
\usepackage{wrapfig}
\usepackage[pdftex]{color}

\begin{document}

% Title and authors
\title{Frequency Multiplexing for Readout of Spin Qubits}
\author{J. M. Hornibrook}
\affiliation{ARC Centre of Excellence for Engineered Quantum Systems, School of Physics, University of Sydney, Sydney, NSW 2006, Australia.}
\author{J. I. Colless}
\affiliation{ARC Centre of Excellence for Engineered Quantum Systems, School of Physics, University of Sydney, Sydney, NSW 2006, Australia.}
\author{A. C. Mahoney}
\affiliation{ARC Centre of Excellence for Engineered Quantum Systems, School of Physics, University of Sydney, Sydney, NSW 2006, Australia.}
\author{X. G. Croot}
\affiliation{ARC Centre of Excellence for Engineered Quantum Systems, School of Physics, University of Sydney, Sydney, NSW 2006, Australia.}
\author{S. Blanvillain}
\affiliation{ARC Centre of Excellence for Engineered Quantum Systems, School of Physics, University of Sydney, Sydney, NSW 2006, Australia.}
\author{H. Lu}
\affiliation{Materials Department, University of California, Santa Barbara, California 93106, USA.}
\author{A. C. Gossard}
\affiliation{Materials Department, University of California, Santa Barbara, California 93106, USA.}
\author{D. J. Reilly$^\dagger$}
\affiliation{ARC Centre of Excellence for Engineered Quantum Systems, School of Physics, University of Sydney, Sydney, NSW 2006, Australia.}

\begin{abstract}
We demonstrate a low loss, chip-level frequency multiplexing scheme for readout of scaled-up spin qubit devices. By integrating separate bias tees and resonator circuits on-chip for each readout channel, we realise dispersive gate-sensing in combination with charge detection based on two rf quantum point contacts (rf-QPCs). We apply this approach to perform multiplexed readout of a double quantum dot in the few-electron regime,  and further demonstrate operation of a 10-channel multiplexing device. Limitations for scaling spin qubit readout to large numbers of multiplexed channels is discussed.
\end{abstract}
% Date and title
\date{\today}
\maketitle

Scaling-up quantum systems to the extent needed for fault-tolerant operation introduces new challenges not apparent in the operation of single or few-qubit devices.  Spin qubits based on gate-defined quantum dots \cite{Hanson:2007eg} are, in principle, scalable, firstly because of their small (sub-micron) footprint, and secondly, since spins are largely immune to electrical disturbance, they exhibit low crosstalk when densely integrated \cite{Science_review}. At the few-qubit level, readout of spin-states is via quantum point contact (QPC) or single electron transistor (SET) charge sensors, proximal to each quantum dot  \cite{DiCarlo,Rimberg,Elzerman:803014,Reilly:2007ig,Amasha:2008ky,Barthel:2009hx}. These readout sensors pose a significant challenge to scale-up however, in that they require separate surface gates and large contact leads, crowding the device and tightly constraining the on-chip architecture. 

The recently developed technique of dispersive gate-sensing (DGS) overcomes this scaling limitation by making use of the gates, already in place to define the quantum dots, as additional charge sensors \cite{colless13}. The gates act as readout detectors by sensing small changes in the quantum capacitance associated with the tunnelling of single electrons. In turn, shifts in capacitance are measured by the  response of a radio-frequency (rf) $LC$ resonator that includes the gate. In principle, all of the quantum dot gates used for electron confinement can also be used as dispersive sensors, simultaneously collecting more of the readout signal that is spread over the total device capacitance and thus increasing the signal to noise ratio. Enabling all-gate readout, as well as multichannel rf-QPC or rf-SET charge sensing, requires the development of multiplexing schemes that scale to large numbers of readout sensors and qubits.

Here we report an on-chip approach to frequency multiplexing for the simultaneous readout of scaled-up spin qubit devices. We demonstrate 3-channel readout of a few-electron double quantum dot, combining two rf-QPCs and a dispersive gate-sensor as well as the operation of a 10-channel planar multiplexing (MUX) circuit. Similar approaches to frequency multiplexing have been demonstrated for distributed resonators in the context of kinetic inductance detectors \cite{day03}, superconducting qubits \cite{chen12,Jerger} and rf-SETs \cite{Stevenson,Buehler,BiercukPRB}. The present work advances previous demonstrations by lithographically integrating the feed-lines, bias tees, and resonators, which are fabricated on a sapphire chip using low-loss superconducting niobium. By putting these components on-chip, the size of the entire MUX circuit is reduced far below the wavelength of the rf signals, suppressing impedance mismatch from the unintentional formation of stub-networks \cite{Pozar} that otherwise occur in macroscale multi-channel feedlines. Finally, we briefly discuss the ultimate limitations to scaling frequency multiplexing for spin qubit readout.

Our readout scheme (Fig. 1(a)) comprises a multiplexing chip fabricated from a single layer of superconducting niobium film (150 nm, $J_c=$ 15 MAcm$^{-2}$, $T_c=$ 8.4 K) on a sapphire substrate (r-cut, 3 mm $\times$ 5 mm $\times$ 0.5 mm) using optical photolithography and argon ion beam milling. The niobium remains superconducting at the moderate magnetic fields needed to operate spin qubits. Each inductor $L_i$ in resonance with the parasitic capacitance $C_p$ defines a unique frequency channel $f_i = 1/(2\pi \sqrt{L_iC_p})$ for addressing each readout detector. This multiplexing chip is mounted proximal to the spin qubit chip, consisting of a GaAs/Al$_{0.3}$Ga$_{0.7}$As heterostructure with two dimensional electron gas (2DEG) 110 nm below the surface (carrier density 2.4 $\times$ 10$^{15}$ m$^{-2}$, mobility 44 m$^2$/V s). Ti/Au surface gates define the quantum dots and readout sensors. Bondwires connect the inductors $L_i$ on the multiplex chip to rf-QPCs  via an ohmic contact \cite{Reilly:2007ig} or directly to the gates for the DGS readout \cite{colless13}. The labels (i) - (iii) in Fig. 1 (b) are used to identify frequency channels for the separate readout detectors. Each resonant circuit contains an integrated bias tee for independent dc voltage biasing. Both the multiplexing chip and qubit chip are housed together in a custom printed circuit board platform \cite{colless_RSI} mounted at the mixing chamber stage of a dilution refrigerator with base temperature 20 mK. 

The on-chip bias tees are constructed using inter-digitated capacitors (Fig. 1(d)) with critical dimension 3 $\mu$m and have size-dependent values between 3 pF and 5 pF, with lower frequency channels requiring a larger capacitance for similar insertion loss. To further increase the coupling capacitance we spin-coat the interdigitated sections with photoresist (AZ6612, $\epsilon \approx 4$) to yield a larger dielectric constant than free space. The inductors (red, Fig. 1(e)), used in both the resonant circuit and bias tees, are spiral shaped with critical dimension 3 $\mu$m. The measured inductances (170, 250 and 400 nH) are in agreement with analytical calculations based on their geometry \cite{mohan99}. The self-resonance frequency of all the inductors is increased by over-etching the sapphire dielectric between adjacent turns, decreasing the effective dielectric constant and reducing the capacitance.  Measurements of the transmitted power for the individual planar components are shown in Fig. 1(f,g) (blue, red trace) and yield agreement with numerics based on a 3D electromagnetic field simulation (black trace) \cite{HFSSQ3D}.
% Figure 1
\begin{figure}
\includegraphics[scale=0.4]{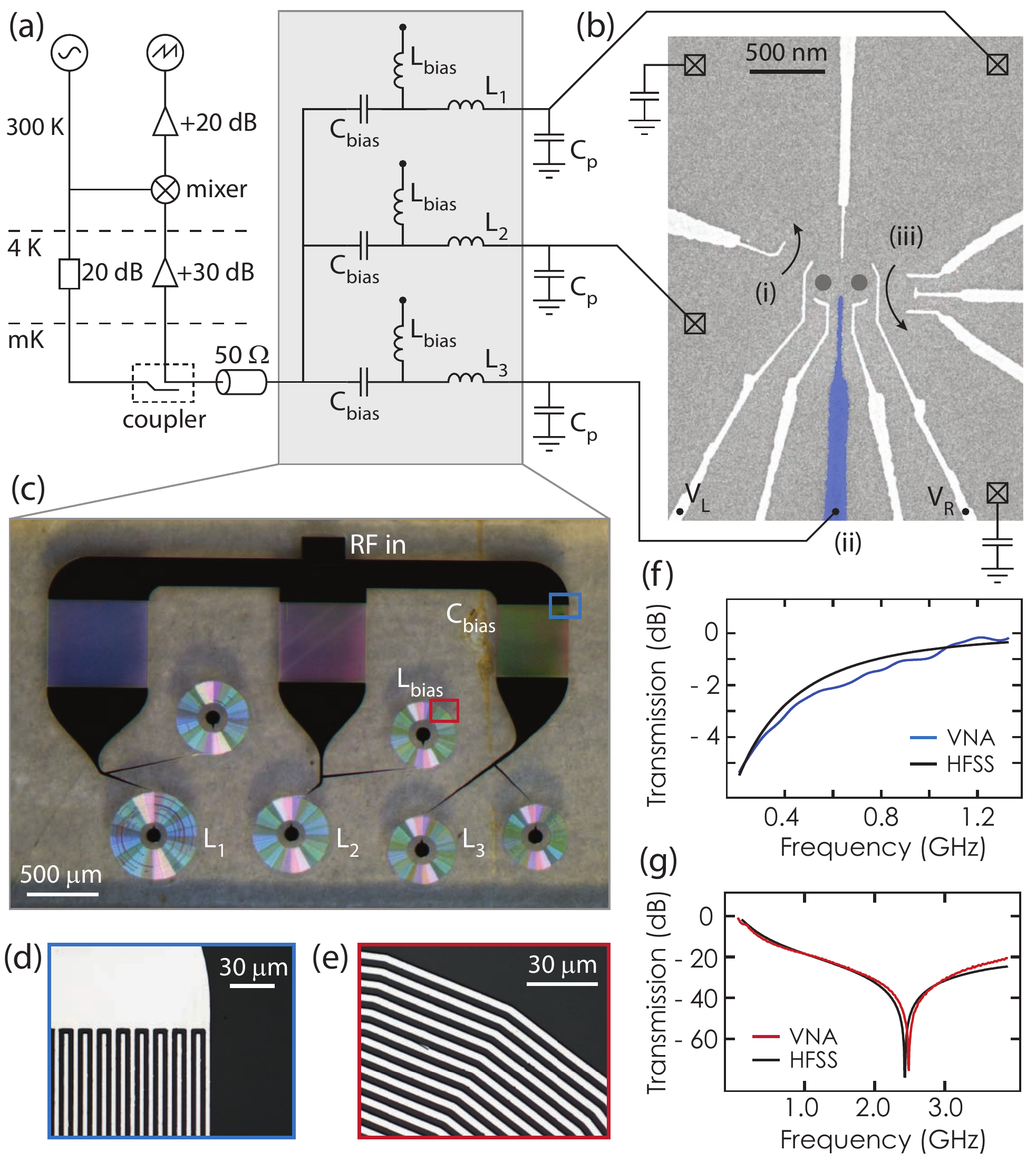}
\caption{\small {\bf(a)} Three channel frequency multiplexing scheme for spin qubit readout. The individual $LC$ resonator circuits comprise a matching inductor $L_i$, parasitic capacitance $C_p$ and a bias tee for independent biasing of each gate sensor. {\bf(b)} Micrograph of the GaAs double dot device. Individual channels of the multiplexing chip are connected via bondwires to either a  gate sensor (labelled (ii)) or an ohmic contact on one side of a QPC (labelled (i), (iii)). {\bf(c)} Optical micrograph of the multiplexing chip which is patterned using niobium on a sapphire substrate, comprising interdigitated capacitors {\bf(d)} and spiral inductors {\bf(e)}. {\bf(f), (g)} Microwave transmission through bias tee components - measurement via a vector network analyser (VNA) and 3D numerical simulation, {\it Ansoft} HFSS \cite{HFSSQ3D}.} 
\end{figure}

% Figure 2
\begin{figure}
\includegraphics[scale=0.4]{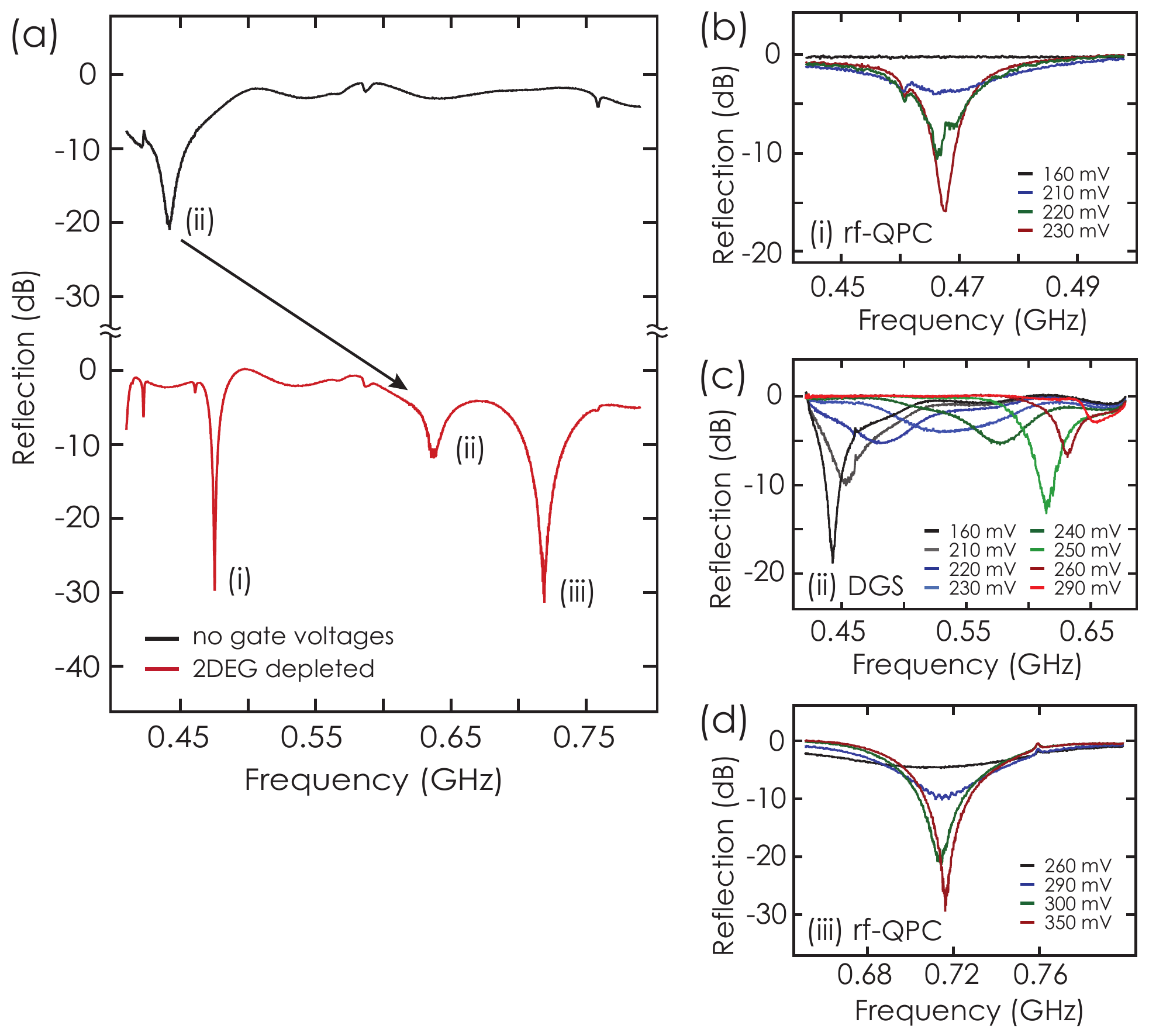}
\caption{\small {\bf (a)} Frequency response of MUX circuit separating left rf-QPCs (i), dispersive gate sensor (ii), and right rf-QPC (iii) into separate frequency channels. With negative voltage applied to the gates, the frequency response (shown in red) exhibits resonances as the impedance of the readout sensors approach the characteristic impedance of the feedline. {\bf (b)} and {\bf (d)} show the frequency response of the left and right rf-QPCs as the gate voltage modulates the conductance. {\bf (c)} shows the frequency response of the dispersive gate sensor with gate bias. Note the significant shift in resonance frequency as the gate capacitance is reduced by depleting the electron gas beneath.}
\end{figure}

The multiplexing scheme is implemented using a 3-channel chip to read out the state of a double quantum dot. The frequency response of the chip strongly depends on the state of the readout detectors, as shown in Fig. 2(a). In the absence of gate bias (black trace), the QPCs are far from pinch-off and the corresponding resonances are not apparent since the impedance of the $LCR$ network is well away from the characteristic impedance of the feedline ($Z_0 \sim 50$ $\Omega$). The resonances are formed (red trace) with the application of negative gate bias, depleting the electron gas and increasing the resistance of the QPC to bring the combined $LCR$ network towards a matched load. Larger gate bias subsequently pinches-off the rf-QPC, further modulating the amount of reflected rf power at the resonance frequency. The response of the gate-sensor with bias is significantly different to that of the rf-QPC. For the gate-sensor, depleting the 2DEG beneath the gate also increases its resonance frequency, as shown in Fig. 2(c). This frequency dependence arises from the change in parasitic capacitance as the electron gas is depleted. With the gate voltages typically needed for defining quantum dots, the parasitic capacitance $C_p$ is of the order of 0.3 pF. Electromagnetic field simulation suggests contributions to $C_p$ are roughly equal between 2DEG, bondwires and adjacent turns of the planar inductors. Given the large separation in resonance frequencies, crosstalk is negligible in this 3-channel implementation.

% Figure 3
\begin{figure}[t]
\includegraphics[scale=0.45]{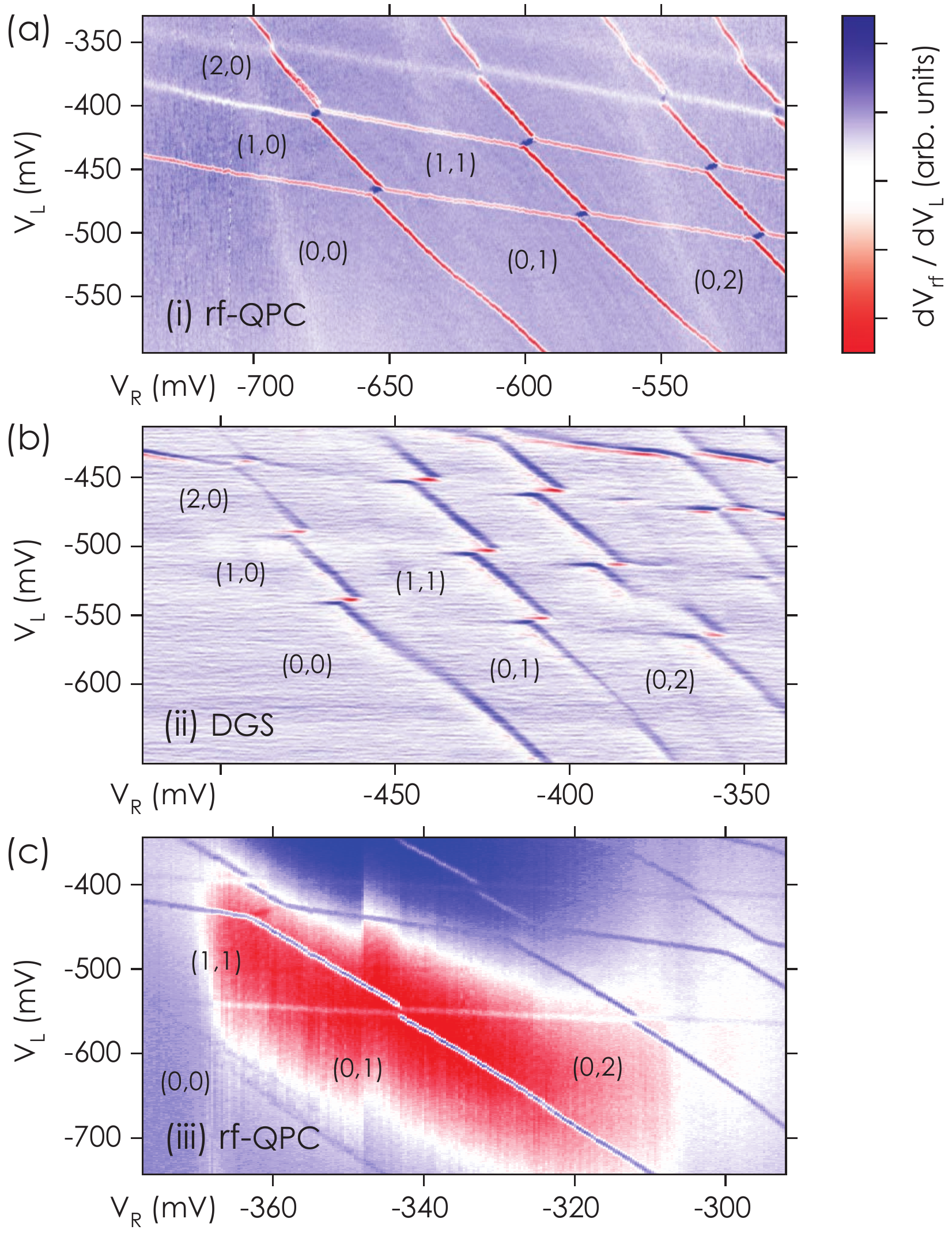}
\caption{\small Multiplexed readout of double quantum dot in the few-electron regime. The derivative of $V_{rf}$ with respect to $V_L$, in arbitrary units, is shown as a function of the voltages on the left and right gates, $V_L$ and $V_R$. Charge stability diagrams {\bf(a), (b), (c)} correspond respectively to readout using the separate channels (i), (ii) and (iii) as indicated in Fig. 2. Electron occupancy in the left and right dots is indicated by the labels (m,n). Note that when biasing the left and right QPC gates (needed for (a) and (c)) a different gate bias $V_L$ and $V_R$ is required for the same electron number.}
\end{figure}
% Figure 4
\begin{figure}[t]
\includegraphics[scale=0.42]{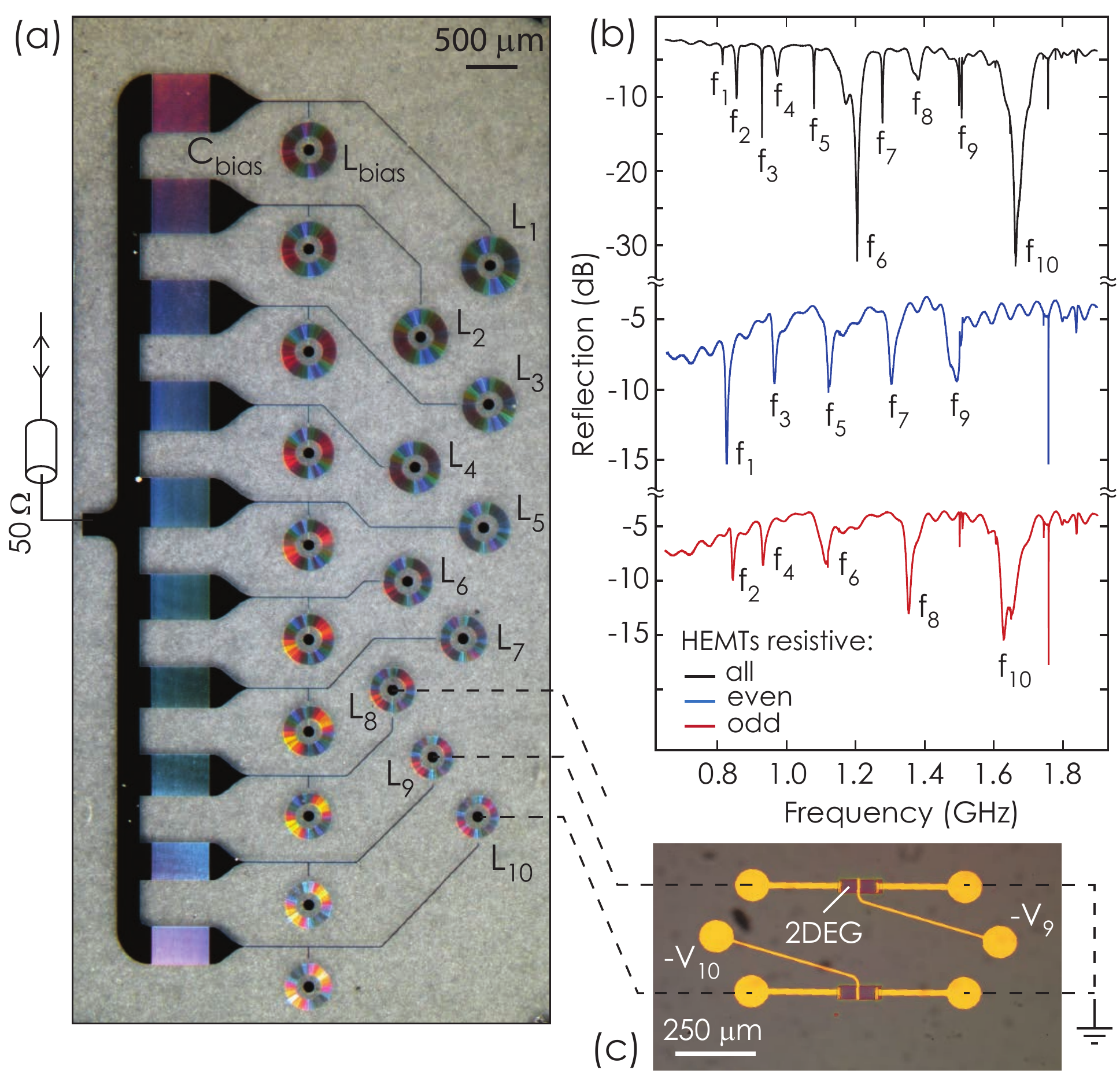}
\caption{\small {\bf (a)} Optical micrograph of a 10:1 MUX chip with integrated bias tees. {\bf (b)} Frequency response of the MUX chip with inductors $L_i$ each connected to GaAs HEMT with a gate-controllable conductance to mimic the response of 10 QPCs. Data shows the response with all HEMTs in the resistive state (black), odd HEMTs resistive  (blue) and even HEMTs resistive (red). {\bf (c)} Shows an optical micrograph of a section of the HEMT device with dashed lines indicative of bondwires.}
\end{figure}

We now demonstrate charge sensing measurements of a double quantum dot in the few-electron regime using this MUX configuration.  The three independent readout channels (i, ii, iii) are separately addressable by selecting the rf carrier to match the respective resonance frequency. We note that direct digital synthesis  can be used to create a single waveform that contains all of the separate carrier frequency components for each channel. The rf signal reflected from the MUX chip is first amplified at cryogenic temperatures before demodulation by mixing the generated and reflected rf tones. Low-pass filtering removes the sum component and, after further baseband amplification, yields a  voltage $V_{rf}$ proportional to the response of the resonance circuit \cite{Reilly:2007ig}. Alternatively, high bandwidth analog to digital conversion can dispense with the need for separate mixers for each channel by directly acquiring the reflected waveform and performing demodulation in software. 

Readout via the QPCs (i and iii) exhibits a typical charge stability diagram in the few-electron regime as a function of gate bias $V_L$ and $V_R$ as shown in Fig. 3(a,c). The label (m,n) denotes the number of electrons in the left and right quantum dot respectively with the colour axis proportional to the derivative of the readout signal with gate bias. In comparison to the rf-QPCs, the dispersive gate readout channel is insensitive to charge transitions that occur with tunnel rates below the resonator frequency \cite{colless13}. Note that biasing the gates to tune the QPCs also shifts the voltages $V_L$ and $V_R$, such that their values are dependent on which sensor is being read out. 

Having demonstrated our approach to frequency multiplexing, we investigate the scalability of this scheme by operating a 10-channel  chip shown in Fig. 4(a). The 10-channels are defined using inductors $L_i$ with values between 60 nH and 250 nH that form a resonant circuit with parasitic capacitance $C_p$ as described above. Each channel again integrates a bias tee, needed for independent biasing of the gate sensors. Operation of the 10-channel chip is tested at 4.2 K using a series of high electron mobility transistors (HEMTs) fabricated from a GaAs/Al$_{0.3}$Ga$_{0.7}$As heterostructure and connected to the MUX chip via bondwires. These HEMTs, shown in Fig. 4(c), act as independent variable resistors and mimic the response of 10 different QPCs for the purpose of testing the MUX scheme. With each HEMT connected to its corresponding resonator, the frequency response of the chip is shown in Fig. 4(b), firstly with all HEMTs in the high resistance state (black trace). Selectivity of each frequency channel is demonstrated by alternatively biasing even-numbered (blue trace) and then odd-numbered (red trace) HEMTs. The exact resonance frequency is set by the contribution to the parasitic capacitance from the HEMT, which depends on the extent to which it is depleted. In this demonstration we have not carefully adjusted the resistance of the HEMTs to optimize the $Q$-factor of each resonator.

Frequency multiplexing allows simultaneous readout but requires separate resonator and bias circuits for each readout channel. Although the size of our demonstration devices are large, the use of alternate fabrication methods will likely alleviate any road-block to scaling based on footprint. For instance, the use of multilayer processing for the capacitors $C_{bias}$ can shrink their footprint to $\sim$ 15 $\mu$m $\times$ 15 $\mu$m for similar capacitance. The space occupied by the bias tee inductors $L_{bias}$ can be suppressed by using resistors instead of inductors to achieve high impedance. Reducing the critical dimension of the resonator inductors to $\sim$ 100 nm results in a 55 $\mu$m $\times$ 55 $\mu$m footprint for the largest (400 nH) inductor used here. Taken together, and assuming these superconducting circuits are fabricated on the same GaAs chip as the qubits, these dimensions suggest that thousands of readout channels are feasible in a moderately sized 1 cm $\times$ 1 cm area.

A more serious challenge is frequency crowding arising from the limited bandwidth available using planar lumped element inductors. For a maximum resonance frequency of $\sim$ 5 GHz and given the need to separate channels by several linewidths to suppress crosstalk, the total number of independent gate sensors that can be read out simultaneously is $\sim$ 100. Beyond this number several approaches are possible. These include a brute force method, duplicating the reflectometry circuit, including cryogenic amplifiers for every bank of 100 channels. Alternatively, the available bandwidth can be extended by making use of distributed resonators \cite{Zmuidzinas}, but these typically have larger footprints. Finally, if the constraint of simultaneous readout is relaxed, time domain multiplexing via cryogenic switching elements would allow readout of banks of frequency multiplexed channels to be interleaved in time. Whether qubit readout via such a time sequenced scheme is possible is likely dependent on the details of the particular quantum algorithm being implemented.

This research was supported by the Office of the Director of National Intelligence, Intelligence Advanced Research Projects Activity (IARPA), through the Army Research Office grant W911NF-12-1-0354 and the Australian Research Council Centre of Excellence Scheme (Grant No. EQuS CE110001013). Device fabrication is made possible through fabrication facilities at CSIRO (Lindfield) and the Australian National Fabrication Facility (ANFF). J. M. H. acknowledges a CSIRO student scholarship.\\

$\dagger$ david.reilly@sydney.edu.au \\

% Bibliography
\bibliographystyle{apsrev4-1}
%\bibliography{Gate-sensingBIB}

\begin{thebibliography}{20}%
\makeatletter
\providecommand \@ifxundefined [1]{%
 \@ifx{#1\undefined}
}%
\providecommand \@ifnum [1]{%
 \ifnum #1\expandafter \@firstoftwo
 \else \expandafter \@secondoftwo
 \fi
}%
\providecommand \@ifx [1]{%
 \ifx #1\expandafter \@firstoftwo
 \else \expandafter \@secondoftwo
 \fi
}%
\providecommand \natexlab [1]{#1}%
\providecommand \enquote  [1]{``#1''}%
\providecommand \bibnamefont  [1]{#1}%
\providecommand \bibfnamefont [1]{#1}%
\providecommand \citenamefont [1]{#1}%
\providecommand \href@noop [0]{\@secondoftwo}%
\providecommand \href [0]{\begingroup \@sanitize@url \@href}%
\providecommand \@href[1]{\@@startlink{#1}\@@href}%
\providecommand \@@href[1]{\endgroup#1\@@endlink}%
\providecommand \@sanitize@url [0]{\catcode `\\12\catcode `\$12\catcode
  `\&12\catcode `\#12\catcode `\^12\catcode `\_12\catcode `\%12\relax}%
\providecommand \@@startlink[1]{}%
\providecommand \@@endlink[0]{}%
\providecommand \url  [0]{\begingroup\@sanitize@url \@url }%
\providecommand \@url [1]{\endgroup\@href {#1}{\urlprefix }}%
\providecommand \urlprefix  [0]{URL }%
\providecommand \Eprint [0]{\href }%
\providecommand \doibase [0]{http://dx.doi.org/}%
\providecommand \selectlanguage [0]{\@gobble}%
\providecommand \bibinfo  [0]{\@secondoftwo}%
\providecommand \bibfield  [0]{\@secondoftwo}%
\providecommand \translation [1]{[#1]}%
\providecommand \BibitemOpen [0]{}%
\providecommand \bibitemStop [0]{}%
\providecommand \bibitemNoStop [0]{.\EOS\space}%
\providecommand \EOS [0]{\spacefactor3000\relax}%
\providecommand \BibitemShut  [1]{\csname bibitem#1\endcsname}%
\let\auto@bib@innerbib\@empty
%</preamble>
\bibitem [{\citenamefont {Hanson}\ \emph {et~al.}(2007)\citenamefont {Hanson},
  \citenamefont {Petta}, \citenamefont {Tarucha},\ and\ \citenamefont
  {Vandersypen}}]{Hanson:2007eg}%
  \BibitemOpen
  \bibfield  {author} {\bibinfo {author} {\bibfnamefont {R.}~\bibnamefont
  {Hanson}}, \bibinfo {author} {\bibfnamefont {J.~R.}\ \bibnamefont {Petta}},
  \bibinfo {author} {\bibfnamefont {S.}~\bibnamefont {Tarucha}}, \ and\
  \bibinfo {author} {\bibfnamefont {L.~M.~K.}\ \bibnamefont {Vandersypen}},\
  }\href@noop {} {\bibfield  {journal} {\bibinfo  {journal} {Rev. Mod. Phys.}\
  }\textbf {\bibinfo {volume} {79}},\ \bibinfo {pages} {1217} (\bibinfo {year}
  {2007})}\BibitemShut {NoStop}%
\bibitem [{\citenamefont {Devoret}\ and\ \citenamefont
  {Schoelkopf}(2013)}]{Science_review}%
  \BibitemOpen
  \bibfield  {author} {\bibinfo {author} {\bibfnamefont {M.~H.}\ \bibnamefont
  {Devoret}}\ and\ \bibinfo {author} {\bibfnamefont {R.~J.}\ \bibnamefont
  {Schoelkopf}},\ }\href@noop {} {\bibfield  {journal} {\bibinfo  {journal}
  {Science}\ }\textbf {\bibinfo {volume} {339}},\ \bibinfo {pages} {1169}
  (\bibinfo {year} {2013})}\BibitemShut {NoStop}%
\bibitem [{\citenamefont {DiCarlo}\ \emph {et~al.}(2004)\citenamefont {DiCarlo}
  \emph {et~al.}}]{DiCarlo}%
  \BibitemOpen
  \bibfield  {author} {\bibinfo {author} {\bibfnamefont {L.}~\bibnamefont
  {DiCarlo}} \emph {et~al.},\ }\href@noop {} {\bibfield  {journal} {\bibinfo
  {journal} {Phys. Rev. Lett.}\ }\textbf {\bibinfo {volume} {92}},\ \bibinfo
  {pages} {226801} (\bibinfo {year} {2004})}\BibitemShut {NoStop}%
\bibitem [{\citenamefont {Lu}\ \emph {et~al.}(2003)\citenamefont {Lu},
  \citenamefont {Ji}, \citenamefont {Pfeiffer}, \citenamefont {West},\ and\
  \citenamefont {Rimberg}}]{Rimberg}%
  \BibitemOpen
  \bibfield  {author} {\bibinfo {author} {\bibfnamefont {W.}~\bibnamefont
  {Lu}}, \bibinfo {author} {\bibfnamefont {Z.}~\bibnamefont {Ji}}, \bibinfo
  {author} {\bibfnamefont {L.~N.}\ \bibnamefont {Pfeiffer}}, \bibinfo {author}
  {\bibfnamefont {K.~W.}\ \bibnamefont {West}}, \ and\ \bibinfo {author}
  {\bibfnamefont {A.~J.}\ \bibnamefont {Rimberg}},\ }\href@noop {} {\bibfield
  {journal} {\bibinfo  {journal} {Nature (London)}\ }\textbf {\bibinfo {volume}
  {423}},\ \bibinfo {pages} {422} (\bibinfo {year} {2003})}\BibitemShut
  {NoStop}%
\bibitem [{\citenamefont {Elzerman}\ \emph {et~al.}(2004)\citenamefont
  {Elzerman}, \citenamefont {Hanson}, \citenamefont {Van~Beveren},
  \citenamefont {Witkamp}, \citenamefont {Vandersypen},\ and\ \citenamefont
  {Kouwenhoven}}]{Elzerman:803014}%
  \BibitemOpen
  \bibfield  {author} {\bibinfo {author} {\bibfnamefont {J.~M.}\ \bibnamefont
  {Elzerman}}, \bibinfo {author} {\bibfnamefont {R.}~\bibnamefont {Hanson}},
  \bibinfo {author} {\bibfnamefont {L.~H.~W.}\ \bibnamefont {Van~Beveren}},
  \bibinfo {author} {\bibfnamefont {B.}~\bibnamefont {Witkamp}}, \bibinfo
  {author} {\bibfnamefont {L.~M.~K.}\ \bibnamefont {Vandersypen}}, \ and\
  \bibinfo {author} {\bibfnamefont {L.~P.}\ \bibnamefont {Kouwenhoven}},\
  }\href@noop {} {\bibfield  {journal} {\bibinfo  {journal} {Nature (London)}\
  }\textbf {\bibinfo {volume} {430}},\ \bibinfo {pages} {431} (\bibinfo {year}
  {2004})}\BibitemShut {NoStop}%
\bibitem [{\citenamefont {Reilly}\ \emph {et~al.}(2007)\citenamefont {Reilly},
  \citenamefont {Marcus}, \citenamefont {Hanson},\ and\ \citenamefont
  {Gossard}}]{Reilly:2007ig}%
  \BibitemOpen
  \bibfield  {author} {\bibinfo {author} {\bibfnamefont {D.~J.}\ \bibnamefont
  {Reilly}}, \bibinfo {author} {\bibfnamefont {C.~M.}\ \bibnamefont {Marcus}},
  \bibinfo {author} {\bibfnamefont {M.~P.}\ \bibnamefont {Hanson}}, \ and\
  \bibinfo {author} {\bibfnamefont {A.~C.}\ \bibnamefont {Gossard}},\
  }\href@noop {} {\bibfield  {journal} {\bibinfo  {journal} {App. Phys. Lett.}\
  }\textbf {\bibinfo {volume} {91}},\ \bibinfo {pages} {162101} (\bibinfo
  {year} {2007})}\BibitemShut {NoStop}%
\bibitem [{\citenamefont {Amasha}\ \emph {et~al.}(2008)\citenamefont {Amasha},
  \citenamefont {MacLean}, \citenamefont {Radu}, \citenamefont {Zumb{\"u}hl},
  \citenamefont {Kastner}, \citenamefont {Hanson},\ and\ \citenamefont
  {Gossard}}]{Amasha:2008ky}%
  \BibitemOpen
  \bibfield  {author} {\bibinfo {author} {\bibfnamefont {S.}~\bibnamefont
  {Amasha}}, \bibinfo {author} {\bibfnamefont {K.}~\bibnamefont {MacLean}},
  \bibinfo {author} {\bibfnamefont {I.}~\bibnamefont {Radu}}, \bibinfo {author}
  {\bibfnamefont {D.}~\bibnamefont {Zumb{\"u}hl}}, \bibinfo {author}
  {\bibfnamefont {M.}~\bibnamefont {Kastner}}, \bibinfo {author} {\bibfnamefont
  {M.}~\bibnamefont {Hanson}}, \ and\ \bibinfo {author} {\bibfnamefont
  {A.}~\bibnamefont {Gossard}},\ }\href@noop {} {\bibfield  {journal} {\bibinfo
   {journal} {Phys. Rev. Lett.}\ }\textbf {\bibinfo {volume} {100}},\ \bibinfo
  {pages} {046803} (\bibinfo {year} {2008})}\BibitemShut {NoStop}%
\bibitem [{\citenamefont {Barthel}\ \emph {et~al.}(2009)\citenamefont
  {Barthel}, \citenamefont {Reilly}, \citenamefont {Marcus}, \citenamefont
  {Hanson},\ and\ \citenamefont {Gossard}}]{Barthel:2009hx}%
  \BibitemOpen
  \bibfield  {author} {\bibinfo {author} {\bibfnamefont {C.}~\bibnamefont
  {Barthel}}, \bibinfo {author} {\bibfnamefont {D.~J.}\ \bibnamefont {Reilly}},
  \bibinfo {author} {\bibfnamefont {C.~M.}\ \bibnamefont {Marcus}}, \bibinfo
  {author} {\bibfnamefont {M.~P.}\ \bibnamefont {Hanson}}, \ and\ \bibinfo
  {author} {\bibfnamefont {A.~C.}\ \bibnamefont {Gossard}},\ }\href@noop {}
  {\bibfield  {journal} {\bibinfo  {journal} {Phys. Rev. Lett.}\ }\textbf
  {\bibinfo {volume} {103}},\ \bibinfo {pages} {160503} (\bibinfo {year}
  {2009})}\BibitemShut {NoStop}%
\bibitem [{\citenamefont {Colless}\ \emph {et~al.}(2013)\citenamefont
  {Colless}, \citenamefont {Mahoney}, \citenamefont {Hornibrook}, \citenamefont
  {Doherty}, \citenamefont {Lu}, \citenamefont {Gossard},\ and\ \citenamefont
  {Reilly}}]{colless13}%
  \BibitemOpen
  \bibfield  {author} {\bibinfo {author} {\bibfnamefont {J.~I.}\ \bibnamefont
  {Colless}}, \bibinfo {author} {\bibfnamefont {A.~C.}\ \bibnamefont
  {Mahoney}}, \bibinfo {author} {\bibfnamefont {J.~M.}\ \bibnamefont
  {Hornibrook}}, \bibinfo {author} {\bibfnamefont {A.~C.}\ \bibnamefont
  {Doherty}}, \bibinfo {author} {\bibfnamefont {H.}~\bibnamefont {Lu}},
  \bibinfo {author} {\bibfnamefont {A.~C.}\ \bibnamefont {Gossard}}, \ and\
  \bibinfo {author} {\bibfnamefont {D.~J.}\ \bibnamefont {Reilly}},\ }\href
  {\doibase 10.1103/PhysRevLett.110.046805} {\bibfield  {journal} {\bibinfo
  {journal} {Phys. Rev. Lett.}\ }\textbf {\bibinfo {volume} {110}},\ \bibinfo
  {pages} {046805} (\bibinfo {year} {2013})}\BibitemShut {NoStop}%
\bibitem [{\citenamefont {Day}\ \emph {et~al.}(2003)\citenamefont {Day},
  \citenamefont {LeDuc}, \citenamefont {Mazin}, \citenamefont {Vayonakis},\
  and\ \citenamefont {Zmuidzinas}}]{day03}%
  \BibitemOpen
  \bibfield  {author} {\bibinfo {author} {\bibfnamefont {P.~K.}\ \bibnamefont
  {Day}}, \bibinfo {author} {\bibfnamefont {H.~G.}\ \bibnamefont {LeDuc}},
  \bibinfo {author} {\bibfnamefont {B.~A.}\ \bibnamefont {Mazin}}, \bibinfo
  {author} {\bibfnamefont {A.}~\bibnamefont {Vayonakis}}, \ and\ \bibinfo
  {author} {\bibfnamefont {J.}~\bibnamefont {Zmuidzinas}},\ }\href@noop {}
  {\bibfield  {journal} {\bibinfo  {journal} {Nature}\ }\textbf {\bibinfo
  {volume} {425}},\ \bibinfo {pages} {817} (\bibinfo {year}
  {2003})}\BibitemShut {NoStop}%
\bibitem [{\citenamefont {{Chen}}\ \emph {et~al.}(2012)\citenamefont {{Chen}}
  \emph {et~al.}}]{chen12}%
  \BibitemOpen
  \bibfield  {author} {\bibinfo {author} {\bibfnamefont {Y.}~\bibnamefont
  {{Chen}}} \emph {et~al.},\ }\href {\doibase 10.1063/1.4764940} {\bibfield
  {journal} {\bibinfo  {journal} {App. Phys. Lett.}\ }\textbf {\bibinfo
  {volume} {101}},\ \bibinfo {eid} {182601} (\bibinfo {year}
  {2012})}\BibitemShut {NoStop}%
\bibitem [{\citenamefont {Jerger}\ \emph {et~al.}(2012)\citenamefont {Jerger}
  \emph {et~al.}}]{Jerger}%
  \BibitemOpen
  \bibfield  {author} {\bibinfo {author} {\bibfnamefont {M.}~\bibnamefont
  {Jerger}} \emph {et~al.},\ }\href@noop {} {\bibfield  {journal} {\bibinfo
  {journal} {App. Phys. Lett.}\ }\textbf {\bibinfo {volume} {101}},\ \bibinfo
  {pages} {042604} (\bibinfo {year} {2012})}\BibitemShut {NoStop}%
\bibitem [{\citenamefont {Stevenson}\ \emph {et~al.}(2002)\citenamefont
  {Stevenson}, \citenamefont {Pellerano}, \citenamefont {Stahle}, \citenamefont
  {Aidala},\ and\ \citenamefont {Schoelkopf}}]{Stevenson}%
  \BibitemOpen
  \bibfield  {author} {\bibinfo {author} {\bibfnamefont {T.}~\bibnamefont
  {Stevenson}}, \bibinfo {author} {\bibfnamefont {F.}~\bibnamefont
  {Pellerano}}, \bibinfo {author} {\bibfnamefont {C.}~\bibnamefont {Stahle}},
  \bibinfo {author} {\bibfnamefont {K.}~\bibnamefont {Aidala}}, \ and\ \bibinfo
  {author} {\bibfnamefont {R.}~\bibnamefont {Schoelkopf}},\ }\href@noop {}
  {\bibfield  {journal} {\bibinfo  {journal} {App. Phys. Lett.}\ }\textbf
  {\bibinfo {volume} {80}},\ \bibinfo {pages} {3012} (\bibinfo {year}
  {2002})}\BibitemShut {NoStop}%
\bibitem [{\citenamefont {Buehler}\ \emph {et~al.}(2005)\citenamefont {Buehler}
  \emph {et~al.}}]{Buehler}%
  \BibitemOpen
  \bibfield  {author} {\bibinfo {author} {\bibfnamefont {T.~M.}\ \bibnamefont
  {Buehler}} \emph {et~al.},\ }\href@noop {} {\bibfield  {journal} {\bibinfo
  {journal} {App. Phys. Lett.}\ }\textbf {\bibinfo {volume} {86}},\ \bibinfo
  {pages} {143117} (\bibinfo {year} {2005})}\BibitemShut {NoStop}%
\bibitem [{\citenamefont {Biercuk}\ \emph {et~al.}(2006)\citenamefont {Biercuk}
  \emph {et~al.}}]{BiercukPRB}%
  \BibitemOpen
  \bibfield  {author} {\bibinfo {author} {\bibfnamefont {M.~J.}\ \bibnamefont
  {Biercuk}} \emph {et~al.},\ }\href@noop {} {\bibfield  {journal} {\bibinfo
  {journal} {Phys. Rev. B.}\ }\textbf {\bibinfo {volume} {73}},\ \bibinfo
  {pages} {201402} (\bibinfo {year} {2006})}\BibitemShut {NoStop}%
\bibitem [{\citenamefont {Pozar}(2012)}]{Pozar}%
  \BibitemOpen
  \bibfield  {author} {\bibinfo {author} {\bibfnamefont {D.~M.}\ \bibnamefont
  {Pozar}},\ }\href@noop {} {\bibfield  {journal} {\bibinfo  {journal}
  {Microwave Engineering. --4th Ed. (Wiley)}\ } (\bibinfo {year}
  {2012})}\BibitemShut {NoStop}%
\bibitem [{\citenamefont {Colless}\ and\ \citenamefont
  {Reilly}(2012)}]{colless_RSI}%
  \BibitemOpen
  \bibfield  {author} {\bibinfo {author} {\bibfnamefont {J.~I.}\ \bibnamefont
  {Colless}}\ and\ \bibinfo {author} {\bibfnamefont {D.~J.}\ \bibnamefont
  {Reilly}},\ }\href@noop {} {\bibfield  {journal} {\bibinfo  {journal} {Rev.
  Sci. Instrum.}\ }\textbf {\bibinfo {volume} {83}},\ \bibinfo {pages} {023902}
  (\bibinfo {year} {2012})}\BibitemShut {NoStop}%
\bibitem [{\citenamefont {Mohan}\ \emph {et~al.}(1999)\citenamefont {Mohan},
  \citenamefont {Henderson}, \citenamefont {Boyd},\ and\ \citenamefont
  {Lee}}]{mohan99}%
  \BibitemOpen
  \bibfield  {author} {\bibinfo {author} {\bibfnamefont {S.~S.}\ \bibnamefont
  {Mohan}}, \bibinfo {author} {\bibfnamefont {M.}~\bibnamefont {Henderson}},
  \bibinfo {author} {\bibfnamefont {S.~P.}\ \bibnamefont {Boyd}}, \ and\
  \bibinfo {author} {\bibfnamefont {T.~H.}\ \bibnamefont {Lee}},\ }\href@noop
  {} {\bibfield  {journal} {\bibinfo  {journal} {IEEE Journal of Solid-State
  Circuits,}\,\ \bibinfo {pages} {1419}} (\bibinfo {year} {1999})}\BibitemShut
  {NoStop}%
\bibitem [{HFS()}]{HFSSQ3D}%
  \BibitemOpen
  \href@noop {} {\bibinfo  {journal} {HFSS Ansoft Corp. and Q3D extractor}\
  }\BibitemShut {NoStop}%
\bibitem [{\citenamefont {Zmuidzinas}(2012)}]{Zmuidzinas}%
  \BibitemOpen
\bibfield  {journal} {  }\bibfield  {author} {\bibinfo {author} {\bibfnamefont
  {J.}~\bibnamefont {Zmuidzinas}},\ }\href@noop {} {\bibfield  {journal}
  {\bibinfo  {journal} {Annu. Rev. Condens. Matter Phys.}\ }\textbf {\bibinfo
  {volume} {3}},\ \bibinfo {pages} {169} (\bibinfo {year} {2012})}\BibitemShut
  {NoStop}%
\end{thebibliography}
%merlin.mbs apsrev4-1.bst 2010-07-25 4.21a (PWD, AO, DPC) hacked
%Control: key (0)
%Control: author (72) initials jnrlst
%Control: editor formatted (1) identically to author
%Control: production of article title (-1) disabled
%Control: page (0) single
%Control: year (1) truncated
%Control: production of eprint (0) enabled
%

\end{document}